# Evaporation-induced freezing dynamics of droplets levitated in acoustic field

Misaki Mitsuno[1] (光野 海祥), Xiao Ma[2] (馬 驍), and Koji Hasegawa[3,a] (長谷川 浩司)

[1] *Graduate School of Engineering, Kogakuin University, Tokyo 163-8677, Japan*

[2] *Marine Environment & Engine System Department, National Maritime Research Institute, Tokyo 181-0004, Japan*

[3] *Department of Mechanical Engineering, Kogakuin University, Tokyo 163-8677, Japan*

## ABSTRACT

This paper presents the evaporation-induced freezing dynamics of pure cyclohexane droplets levitated via acoustic levitation. Acoustic levitation has attracted considerable attention across various fields owing to its potential to create lab-in-a-drop systems. While droplet evaporation is a fundamental physicochemical process in such a platform, the freezing of droplets induced by evaporation has been sparsely explored experimentally. For pure cyclohexane, the rapid evaporation of levitated droplets initiated freezing at the droplet surface. To better understand this evaporation-induced freezing process, the evaporation behavior of the levitated cyclohexane droplets was visualized and quantified using a high-speed camera and an infrared camera. According to the obtained experimental data, the evaporative heat transfer characteristics of the droplets were identified with theoretical models. Using the derived heat transfer coefficient, a mathematical prediction method for the onset of freezing was proposed and validated with the experimental data. These experimental findings offer valuable insights into the phase transition process and its potential physicochemical applications in a containerless environment.

[a] Corresponding author: Email: kojihasegawa@cc.kogakuin.ac.jp

## I. INTRODUCTION

Droplet evaporation is a ubiquitous phenomenon observed across various contexts, including everyday life and the natural environment. Additionally, droplet evaporation plays a critical role in several industrial applications, such as inkjet printing technology[1,2] and fuel evaporation in internal combustion engines.[3,4] The evaporation of sessile droplets[5] on substrates has been extensively studied, with a focus on internal flow,[6] wettability[7,8] and heat transport.[9,10] However, using a substrate may affect the initial behavior of the droplet. This highlights the need to observe droplets in a free interfacial state, free from substrate contact. In recent years, non-contact fluid manipulation has garnered considerable attention in fields such as analytical chemistry,[11,12] biology,[13,14] and materials science.[15] There are many methodologies for levitation, including acoustic, magnetic, electrostatic, optical, and aerodynamic.[16] While magnetic, electrostatic, optical, and aerodynamic levitation require specific conditions for sample levitation, acoustic levitation can be used for a wide range of applications with various samples, including magnetic and chargeable ones. Therefore, among the various techniques developed, acoustic levitation stands out as a versatile and innovative approach. This technique can levitate samples with any thermophysical properties in midair through resonant acoustic fields, enabling a wide range of lab-in-a-drop processes, including droplet levitation/transport,[16,17] coalescence,[18] mixing,[19,20] reaction,[21,22] evaporation,[23] and solidification.[24,25] Acoustic levitation offers the advantage of mitigating wall effects such as nucleation and contamination, which are common in conventional container-based methods. Despite its promising applications, the strong acoustic fields required for levitation induce complex, unsteady, and nonlinear effects on levitated droplets, including internal[26]/external[27,28] flow fields, temperature fields,[29] concentration fields,[30] and dynamic interfacial behavior, such as interfacial oscillation,[31] deformation,[32] and atomization.[33,34] These phenomena have been explored in both theoretical and experimental studies owing to their profound impact on the evaporation dynamics of levitated samples. The fundamental physics of evaporation involves the transport of mass and heat from the surface of a liquid to the surrounding gas phase. Single-component droplets provide an ideal platform for studying the fundamental physical mechanisms of this transport phenomenon, even in acoustic levitation. Insights into the flow dynamics around and within levitated droplets have been analyzed both theoretically[35] and experimentally,[36] underscoring the importance of mass transport in evaporation processes. Moreover, numerical simulations[28] and mathematical models[37] have enhanced the understanding of droplet evaporation under the influence of acoustic levitation, emphasizing the critical roles of droplet deformation and acoustic flow.

The evaporation of multicomponent droplets,[38] such as water–ethanol mixtures,[39,40] is a complex phenomenon with a broad range of potential applications and has been extensively analyzed. Among the phenomena observed is the evaporation of tri-component droplets, which gives rise to complex physical behaviors due to the interaction of the individual liquids. Additionally, evaporation-induced emulsification, exemplified by the ouzo effect, has been confirmed and received increasing attention in recent years.[41,42]

Another emerging issue is evaporation-induced phase separation and freezing in levitated droplets.[43,44] In light of these findings, an experimental analysis revealed the viability of multiple phase-change processes, including condensation, phase separation and freezing, occurring in a levitated state within an acoustic field. To fundamentally understand the evaporation-induced phase transition process, the temperature and concentration changes on the droplet during evaporation must be explored. However, the effects of heat and mass transport on the underlying physics of these processes remain unclear.[44] Previous studies have examined evaporation-induced cooling and Marangoni effects in acoustically levitated droplets, often focusing on water–ethanol[40] or water–glycerol[9] systems where multicomponent effects dominate. However, these systems frequently involve complex mutual solubility and hygroscopic behavior, making it difficult to isolate the fundamental heat and mass transport processes under acoustic streaming. In contrast, our study uses pure cyclohexane to eliminate such complexity, enabling direct quantification of the heat transfer coefficient and freezing onset. Furthermore, unlike prior theoretical models assuming idealized heat transfer conditions, our work couples experimental infrared thermography with energy-balance modeling under acoustic confinement, thereby offering a more realistic framework for understanding phase changes in levitated droplets. The objective of this study is to provide a deeper understanding of the evaporation-induced dynamics in acoustic levitation. The evaporation-induced freezing dynamics of acoustically levitated pure cyclohexane droplets were explored. Cyclohexane was selected as a model fluid for the present study owing to its high vapor pressure, low latent heat of evaporation, and immiscibility with water. These characteristics allowed us to isolate the evaporation-induced thermal effects without interference from water condensation or mutual solubility, which are common in alcohol-based systems. Furthermore, its well-defined freezing point facilitates the detection and analysis of phase-transition events. The experimental heat transfer properties of phase change processes on levitated droplets were theoretically validated. The results of this study contribute to a more precise understanding of multicomponent droplet behavior and lay the groundwork for innovative developments in droplet-based technologies.

## II. EXPERIMENTAL SETUP

Figure 1(a) presents a schematic of the experimental setup used in this study.[41] A sinusoidal signal generated by a function generator was supplied to an ultrasonic transducer, which emitted ultrasonic waves from the lower horn. These waves were then reflected by the upper reflector, positioned at a multiple of half the wavelength distance, forming an acoustic standing wave between the horn and reflector, as illustrated in Fig. 1(b). To levitate the droplet, a liquid droplet was manually injected near the acoustic pressure node of the standing wave using a syringe and needle. Cyclohexane and ethanol were selected as test liquids for this study. The reflector used had a radius of curvature of 36 mm (R36). The frequency of the acoustic field was 19.3 kHz, corresponding to a wavelength ($\lambda$) of 18 mm. The width of both the horn and the reflector was 36 mm (=2$\lambda$), and the distance between them was 48 mm (~5$\lambda$/2). The sound pressure in the test section was 1.5 ± 0.3 kPa. Droplets were levitated near the third pressure node (~3$\lambda$/2 from the bottom reflector). Experiments were conducted at an ambient temperature of 20 ± 5 °C. To regulate relative humidity, the acoustic levitator was enclosed within a chamber box equipped with a dehumidifier. The ambient temperature and humidity levels were monitored using a high-precision thermometer and hygrometer (Testo, testo622).

Visualization of the levitated droplets was performed using a high-speed camera (Photron, FASTCAM Mini AX200) with backlight illumination. Simultaneously, the droplet interface temperature was recorded using an infrared (IR) camera (FLIR Systems, A6750sc, MWIR). The high-speed and IR cameras were positioned at a 90° angle relative to each other. This configuration allowed us to simultaneously capture the droplet interface with the high-speed camera and the interfacial temperature of the droplet with the IR camera. The high-speed camera operated at a frame rate of 50 fps with a resolution of 256 × 256 pixels, while the IR camera recorded at 1 fps with a resolution of 640 × 512 pixels. The emissivity of the IR camera was set to 0.96.[45] Images captured by the high-speed camera were processed using a computer and in-house code (MATLAB R2023a) to quantify droplet diameter. The droplet diameters $d$ used in this study were equivalent droplet diameters. The initial equivalent droplet diameters $d_0$ ranged from 1.0 to 2.2 mm. To characterize the freezing dynamics of cyclohexane droplets, a space–time diagram was generated using image-processing techniques with in-house code (Python 3.10.3). A step-by-step explanation of this method is provided in a previous publication.[41]

The uncertainty in droplet diameter was <1.4%, because at $d$ = 1.0 mm (the smallest diameter in the present study), the standard deviation with three measurements was <2 pixels, with a spatial resolution of ~15 μm/pixel. Meanwhile, the uncertainty of the droplet surface temperature was ±2 °C with the IR camera used in the present study.

Table I presents the thermophysical properties of water, ethanol, and cyclohexane. Notably, solubility plays a critical role in the evaporation process of levitated droplets. Ethanol, being soluble in water, readily forms a water–ethanol mixture as water condenses from the surrounding vapor, whereas cyclohexane, which is insoluble in water, remains unaffected by such condensation effects. Among the candidate fluids, cyclohexane offers an ideal platform to study evaporation-induced freezing because it is nonpolar and does not absorb ambient moisture, ensuring that the observed phase transition results solely from evaporative cooling. These features enable a more direct comparison between theoretical predictions and experimental measurements in a controlled acoustic environment. For the droplet levitation in the present study, the estimated acoustic impedance of the droplets ($=\rho_l c$) from Table I was sufficiently large, on the order of $10^5$–$10^6$. Therefore, acoustic waves were almost reflected on the droplet surface.

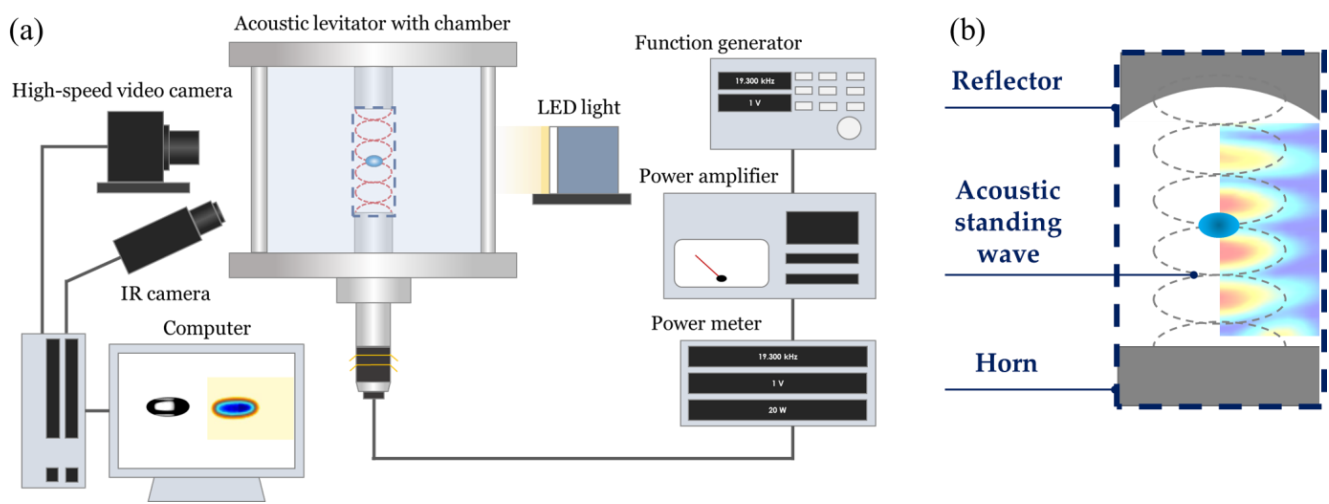

FIG. 1. Schematic of the acoustic levitator: (a) experimental setup and (b) principle of levitation.

TABLE I. Thermophysical properties of the test sample at 25 ºC.

| Fluid | Density $\rho_l$ [kg/m³] | Diffusion coefficient $D \times 10^{-5}$ [m²/s] | Vapor pressure $p_s$ [kPa] | Latent heat of evaporation $L_e$ [kJ/kg] | Specific heat at constant pressure $c_p$ [kJ/kg · K] | Speed of sound $c$ [m/s] | Solubility |
|---|---|---|---|---|---|---|---|
| Water[46] | 997 | 2.49 | 3.17 | 2257 | 4.17 | 1496 | Miscible in ethanol Immiscible in cyclohexane |
| Ethanol[46] | 785 | 1.20 | 7.89 | 855 | 2.42 | 1255 | Miscible in water and cyclohexane |
| Cyclohexane[47] | 774 | 0.59 | 13.02 | 358 | 1.85 | 1141 | Miscible in ethanol Immiscible in water |

## III. RESULTS AND DISCUSSION

### A. Evaporation-induced freezing process of droplets in acoustic levitation

To investigate the evaporation dynamics of acoustically levitated volatile droplets, the temporal variation in droplet size was quantitatively analyzed. Figure 2 illustrates the evaporation process of cyclohexane and ethanol droplets. Cyclohexane and ethanol sample droplets were levitated and measured multiple times under a relative humidity of 40%. Representative data are shown for cyclohexane at $d_0 = 2.0$ mm and ethanol at $d_0 = 1.5$ mm. As shown in Fig. 2(a), the cyclohexane droplet darkened at 26 s, indicating the onset of freezing. Subsequently, the frozen cyclohexane transformed from an ellipsoidal shape to an irregular form. In contrast, the ethanol droplet continuously evaporated without undergoing freezing.

Figure 2(c) presents the time evolution of the evaporating droplets based on the images in Figs. 2(a) and (b). The vertical axis represents the normalized surface area, defined as the droplet surface area $S$ divided by the initial surface area $S_0$.[35] The experimental results confirm that both droplets exhibited a gradual decrease in diameter due to evaporation. Notably, cyclohexane evaporated more rapidly than ethanol because of its higher saturation vapor pressure.

In Fig. 2(c), the solid line represents the theoretical predictions for cyclohexane and ethanol evaporation based on the mathematical model describing single-droplet evaporation. The evaporation model is expressed as follows:[45,48]

$$\frac{S}{S_0} = 1 - \frac{8\pi DM}{\rho_l R}\left(\frac{P_s}{T_s} - \frac{P_\infty}{T_\infty}\right)\frac{t}{S_0}, \tag{1}$$

where $D$ denotes the diffusion coefficient, $M$ denotes the molar mass, $\rho_l$ denotes the liquid density, $R$ denotes the gas constant, $P$ denotes the vapor pressure, $T$ denotes the temperature, and $t$ denotes time. The subscripts $s$ and $\infty$ represent the droplet surface and ambient conditions, respectively.

As shown in Fig. 2(c), the evaporation process of cyclohexane predicted by the evaporation model underestimated the experimental values. This discrepancy is attributed to the increased evaporation rate caused by acoustic streaming around the levitated droplet. The evaporative heat transfer characteristics are further discussed in the following section. Meanwhile, the ethanol droplet exhibited good agreement with the evaporation model for the first 130 s. However, beyond this point, the model prediction began to underestimate the experimental results. According to a previous study,[39] this deviation is attributed to water vapor condensation around the droplet, as ethanol is highly soluble in water.

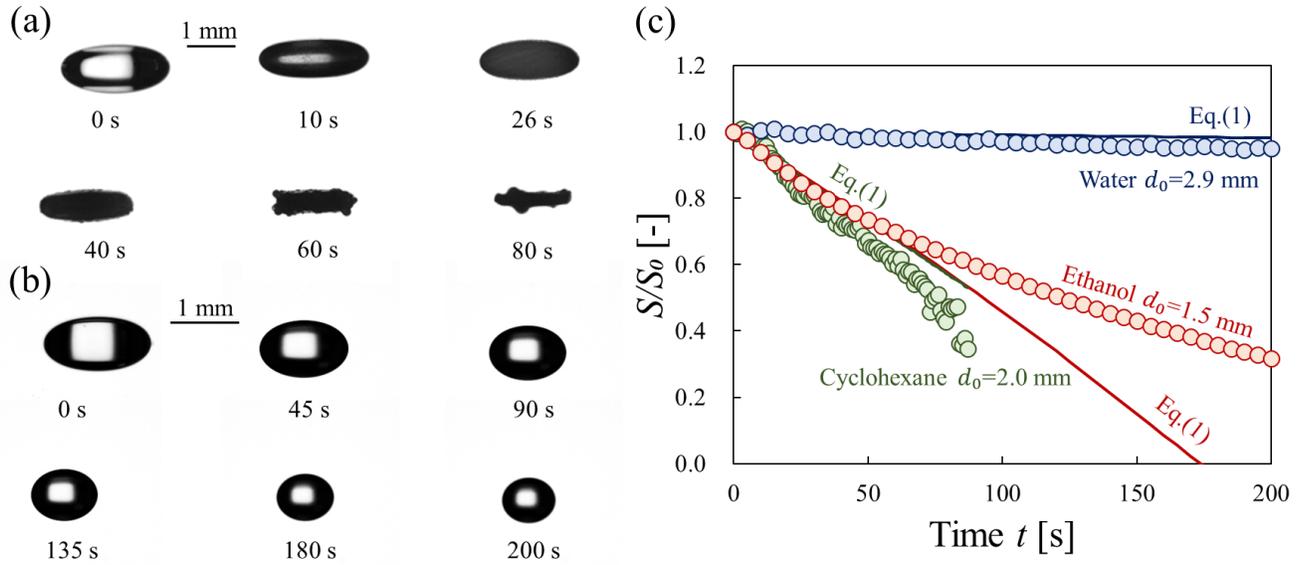

FIG. 2. Evaporation process of levitated pure volatile droplets: (a) snapshots of cyclohexane, (b) snapshots of ethanol, and (c) time evolution of evaporating cyclohexane, ethanol, and water droplets.

Figure 3(a) illustrates the evaporation process and a snapshot of a cyclohexane droplet undergoing freezing while levitated. The inset images correspond to each snapshot shown in Fig. 2(a), where the droplet transitions from an ellipsoid to an irregular shape. This morphological change is attributed to the surface temperature dropping below the freezing point of cyclohexane, leading to solidification. To verify this, Fig. 3(b) presents the surface temperature evolution recorded by an IR camera. The black dotted line represents the ambient temperature (16 °C), while the blue dashed line indicates the freezing point of cyclohexane (6.5 °C). Immediately after levitation, the surface temperature of the droplet rapidly decreased below its freezing point. Subsequently, the temperature increased to the freezing point within 25 s, confirming the onset of freezing. Notably, despite being below the freezing point, the cyclohexane droplet remained in a liquid state because of supercooling, as the temperature declined gradually owing to evaporation. At approximately 25 s, the surface temperature increased owing to heat recovery, suggesting that the droplet had fully frozen. These experimental results demonstrated that cyclohexane droplets levitated in an acoustic field experience supercooling due to evaporative cooling, followed by a phase transition at a critical point. The transition from supercooling to freezing is likely affected by acoustic pressure fluctuations, external flow dynamics, and interfacial motion. Once frozen, the surface temperature of the droplet stabilized, indicating a thermal equilibrium between the latent heat released during freezing and the heat absorbed through evaporation.

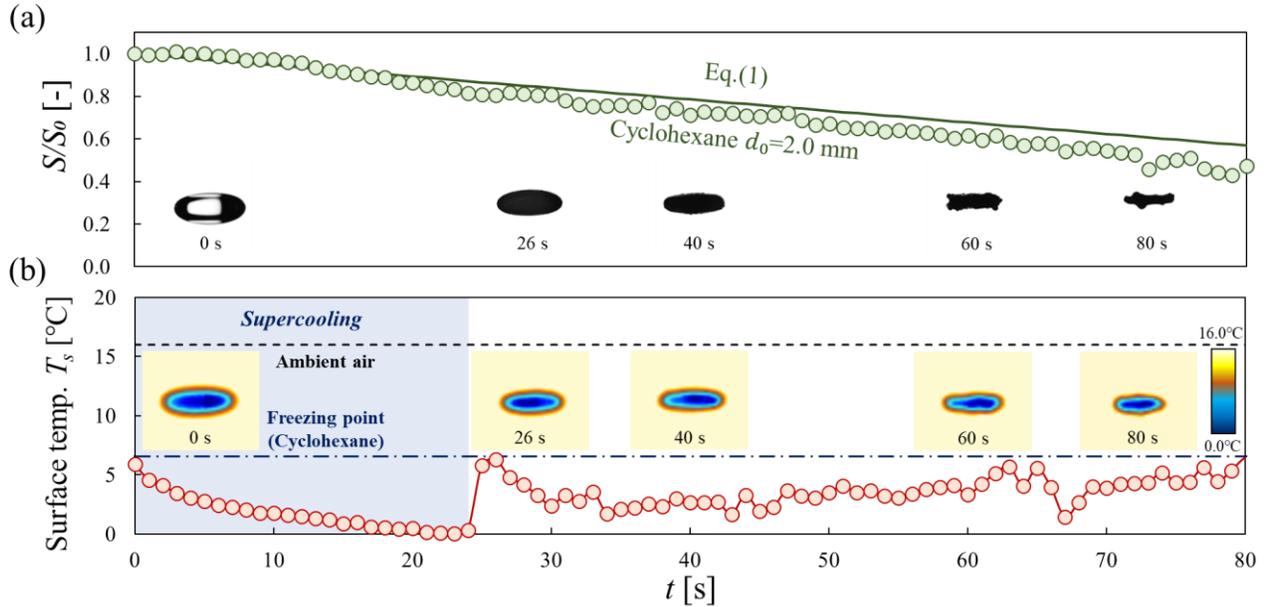

FIG. 3. Evaporation process and surface temperature of a cyclohexane droplet: (a) time evolution of $S/S_0$ and (b) surface temperature variation.

Figure 4 illustrates the freezing process of cyclohexane in an acoustic field. Figure 4(a) presents a top-view snapshot of a levitated droplet with an initial diameter of 2.0 mm at a relative humidity of 45%. The snapshot reveals a freezing front on the droplet's surface immediately after levitation. This observation suggests that the droplet remained in a supercooled state between 0 and 5 s while the freezing front was visible and that complete freezing was achieved at 9 s, marking the transition out of the supercooled state. The propagation of solidification occurred within approximately 1 s, indicating a rapid freezing process following the termination of supercooling. Figure 4(b) presents a space–time diagram based on Fig. 4(a), where the white regions indicate areas of light reflection, signifying the phase transition to a frozen state. The results confirm that the freezing process began partially and progressed to the entire droplet. This phenomenon, which is unique to acoustic levitation, is attributed to the transition from the supercooled state to the frozen state. According to these observations, the solidification process of levitated cyclohexane droplets in an acoustic field can be conceptualized as follows (Fig. 4(c)): (1) reduction of surface temperature due to rapid evaporation and simultaneous water condensation, (2) transition to a supercooled state, (3) onset of freezing, and (4) completion of solidification. Notably, the supercooled state in step (2) is facilitated by the containerless environment provided by acoustic levitation.

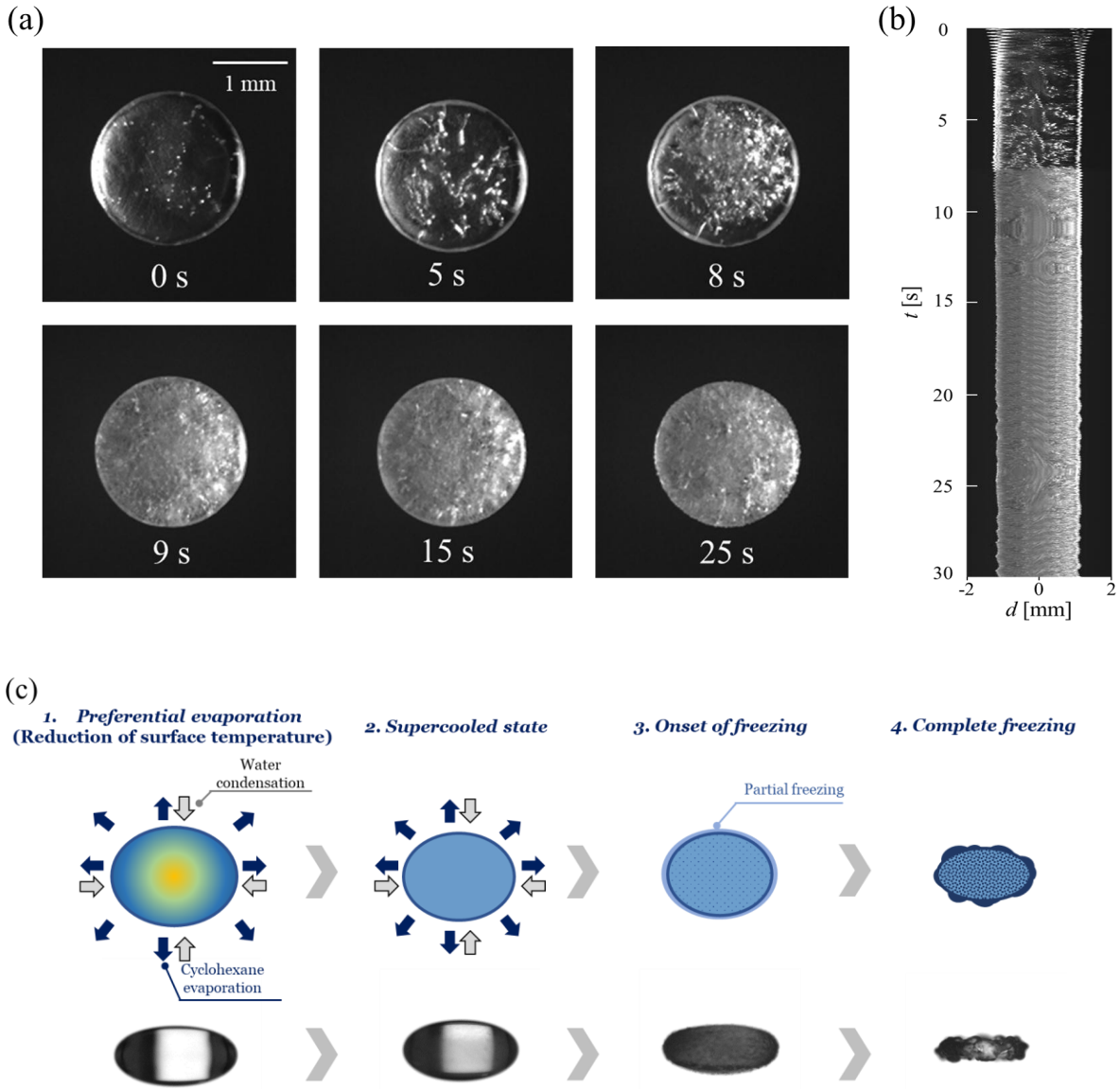

FIG. 4. Evaporation-induced freezing process of a droplet: (a) top-view snapshot of the droplet, (b) space–time diagram of the pure droplet, and (c) schematic of the freezing process in acoustic levitation.

## B. Heat transfer characteristics of cyclohexane droplets

As discussed earlier, this study indicated that cyclohexane droplets undergo a phase change, freezing as they evaporate. Because this phenomenon is primarily governed by heat transfer around the droplet, this section evaluates the heat transfer characteristics of cyclohexane levitated in an acoustic field. The heat transfer coefficient for acoustically levitated droplets can be estimated according to the volume reduction and the temperature gradient near the droplet surface. Assuming that all heat loss occurs solely through evaporation, the heat dissipation due to evaporation can be expressed as follows:

$$Q_{evap} = -L_e \rho_l \frac{dV}{dt}, \qquad (2)$$

$$Q_{ht} = -hS(T_s - T_\infty), \qquad (3)$$

$$Q_{evap} = Q_{ht}, \qquad (4)$$

where $Q$ represents heat quantity. It is divided into $Q_{evap}$, which is the heat quantity used for evaporation, and $Q_{ht}$, which is the heat quantity used for heat transfer. $L_e$ denotes the latent heat of evaporation, $dV/dt$ denotes the evaporation rate of the droplet volume, $T_s$ denotes the droplet surface temperature, and $T_\infty$ denotes the ambient temperature. The evaporation rate ($dV/dt$) for water and cyclohexane was determined from the slope of the linear approximation of the droplet volume variation over time. By combining Eqs. (2)–(4), the heat transfer coefficient can be calculated as follows:

$$h_{exp} = \frac{L_e \rho_l \, dV/dt}{S(T_s - T_\infty)}. \qquad (5)$$

In this study, the droplet diameter $d'$ was defined at the point where the droplet reached its lowest surface temperature, corresponding to the maximum $T_s - T_\infty$. The temperature differences for each droplet, obtained experimentally, are presented in Fig. 5. The results indicated that the temperature differences were almost constant regardless of the droplet diameter. Therefore, the temperature difference of 15.3°C was used for the following analysis.

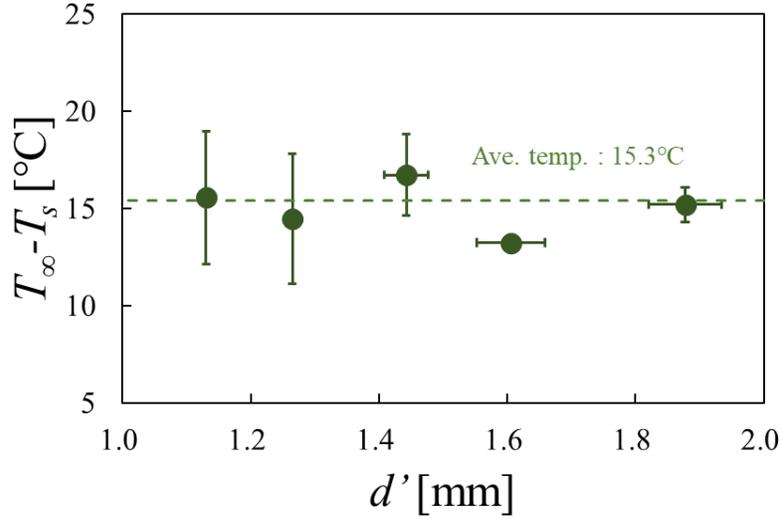

FIG. 5. Maximum temperature difference between the ambient temperature and the cyclohexane surface temperature. Error bars indicate the standard deviation for each plot.

To validate the heat transfer coefficient obtained from Eq. (5) and experimental data, the experimentally derived heat transfer coefficient $h_{exp}$ for cyclohexane droplets was compared with the theoretical prediction in Fig. 6(a). The experimental results indicate that the heat transfer coefficient increases as the droplet diameter decreases. Compared with the model prediction under natural convection conditions ($Nu = 2$), the experimental values exhibit higher heat transfer coefficients, suggesting the influence of forced convection around the droplet. $Nu$ denotes the Nusselt number. The dashed line in the figure represents the heat transfer coefficient $h_{RM}$, which was calculated using the Ranz–Marshall correlation.[49]

$$Nu = 2 + 0.6 Re^{1/2} Pr^{1/3},$$
$$(1 < Re < 10^5, 0.6 < Pr < 380)$$
(6)

$$Re = \frac{\rho_a U_a d}{\mu_a}, Pr = \frac{\mu_a c_p}{k_a}, \qquad (7)$$

$$h_{RM} = \frac{k_a Nu}{d'}, \qquad (8)$$

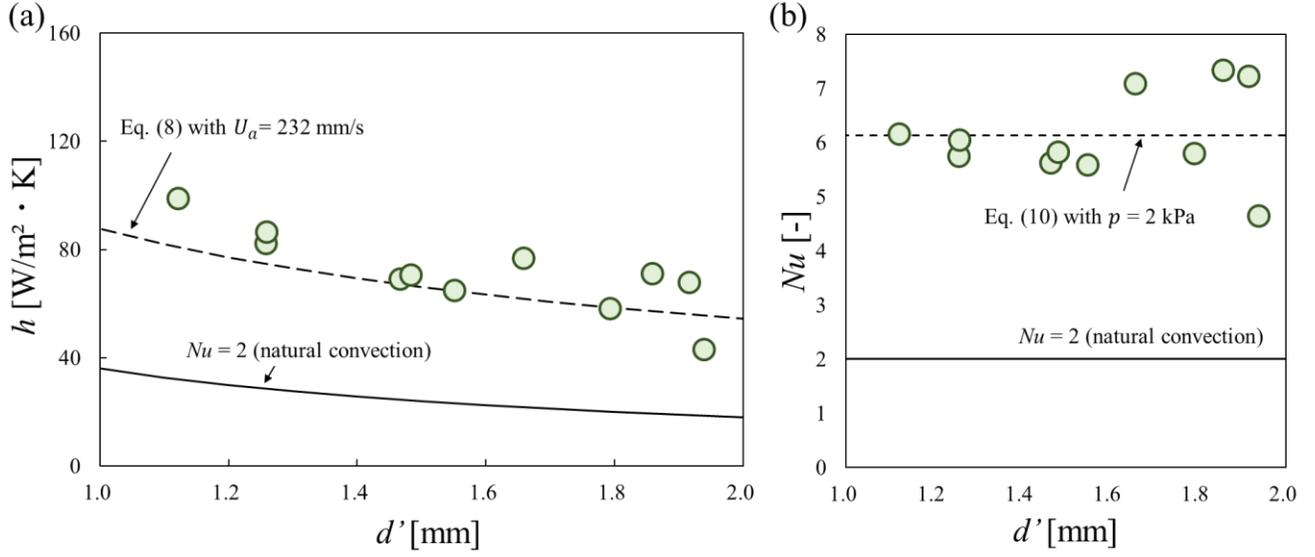

FIG. 6. Heat transfer characteristics of levitated cyclohexane droplets: (a) experimentally derived heat transfer coefficient and (b) Nusselt number.

where $\mu_a$ denotes the air viscosity, $k_a$ denotes the thermal conductivity of air (18 mW/m·K), $c_p$ denotes the specific heat at constant pressure, $U_a$ denotes the velocity of the external flow, $Re$ denotes the Reynolds number, and $Pr$ denotes the Prandtl number. The acoustic streaming velocity of cyclohexane ($U_a$) was measured to be 232 ± 31 mm/s in this study. The velocity was determined by dividing the displacement of the water mist produced by the nebulizer by the time difference. The measurement method used was similar to those employed in previous studies,[50] and the average value was calculated from 10 sets of measurements. The calculated heat transfer coefficient $h_{RM}$ obtained from the Ranz–Marshall correlation agreed well with the heat transfer coefficient $h_{exp}$ obtained from the experimental results. For smaller droplets, some data indicated that the experimental results exceeded the theoretical predictions.

This slight discrepancy is attributed to the disparity between the flow field surrounding the droplet as postulated in the Ranz–Marshall equation and the flow field surrounding the levitated droplet in the present study. The heat transfer coefficient derived from the Ranz–Marshall equation assumes heat transfer for a sphere in a uniform flow field; however, circulating vortices have been observed near the interface of levitated droplets in an acoustic field[51]. Owing to the complex flow structure around the droplet, it can be considered that the heat transfer coefficients obtained in this experiment were slightly higher than those predicted by the Ranz–Marshall equation.

To characterize the impact of acoustic streaming on the heat transfer characteristics of levitated droplets, the Nusselt number was obtained from the experimental heat transfer coefficient. Figure 6(b) shows the Nusselt number calculated from the experimental results using the following equation:

$$Nu = \frac{h_{exp} d'}{k_a}. \tag{9}$$

The solid line represents natural convection conditions with $Nu = 2$, and the dashed line indicates the average $Nu$ in the acoustic boundary-layer flow near the surface of the quasi-spherical droplet at $p = 2$ kPa, as given by Eq. (10).[35]

$$\overline{Nu} = 1.89 \frac{p}{c_0 \rho_a \sqrt{\omega \alpha_a}} \tag{10}$$

Here, $p$ denotes the sound pressure, $c_0$ denotes the speed of sound in air, $\rho_a$ denotes the air density, $\omega$ denotes the angular frequency, and $\alpha_a$ denotes the thermal diffusivity of air. As illustrated in Fig. 6(b), the experimental results demonstrate that levitated cyclohexane in an acoustic field exhibits a higher $Nu$ than under natural convection conditions. This phenomenon is attributed to the external flow occurring within the droplet, which enhances heat transfer, increasing $Nu$. The theoretical model, depicted by the dashed line in Fig. 6(b), provides a satisfactory prediction of the experimental outcomes. This observation signifies that the presence of cyclohexane droplets facilitates heat transfer through the mechanism of boundary-layer acoustic streaming. The experimental findings indicate that cyclohexane levitated within an acoustic field exhibits enhanced heat transfer performance. This heat transfer enhancement is attributed to two primary factors: the high saturation vapor pressure of cyclohexane and the effects of boundary-layer acoustic streaming. These good heat transfer characteristics are essential for freezing cyclohexane droplets in acoustic fields.

## C. Onset of freezing time predicted by model

This subsection presents a mathematical model for predicting the onset of freezing time of a levitated cyclohexane droplet in an acoustic field. The freezing process can be modeled according to the energy balance between the cooling process due to sensible heat transport within the droplet ($Q_{in}$) and the heat transport mechanisms outside the droplet, including heat transfer, heat conduction, and latent heat transport ($Q_{out}$). The thermal energy components can be calculated using the following equations:

$$Q_{in} = \rho_l \frac{dV}{dt} c_p (T_s - T_\infty), \qquad (11)$$

$$Q_{out} = -hS'(T_s - T_\infty) - k_a S' \frac{(T_s - T_\infty)}{\delta_T} - L_e \rho_l \frac{dV}{dt}. \qquad (12)$$

During the freezing of a levitated droplet in a sound field, the energy balance can be expressed as

$$\rho_l \frac{dV}{dt} c_p (T_s - T_\infty) = -hS'(T_s - T_\infty) - k_a S' \frac{(T_s - T_\infty)}{\delta_T} - L_e \rho_l \frac{dV}{dt}, \qquad (13)$$

where $L_e$ denotes the latent heat of evaporation, and $\delta_T (\sim k_a/h)$ represents the thickness of the thermal boundary layer. The temperature difference $T_s - T_\infty$ was taken as the average value of 15.3 °C for cyclohexane, as obtained from Fig. 5. The droplet volume $V'(= \pi d'^3/6)$ and surface area $S'(= \pi d'^2)$ were determined using the experimental data. The onset of freezing time ($t_{fr}$) for levitated droplets can be estimated using the following equation:

$$t_{fr} = \frac{\rho_l V' \left(L_e + c_p(T_s - T_\infty)\right)}{hS'(T_s - T_\infty) + k_a S' \frac{(T_s - T_\infty)}{\delta_T}}. \tag{14}$$

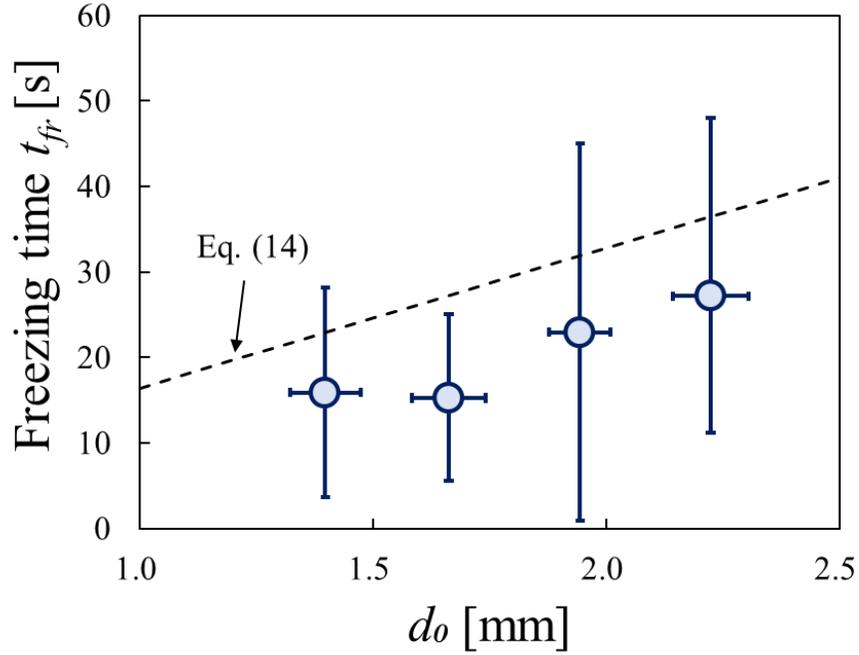

FIG. 7. Onset of freezing time for acoustically levitated cyclohexane droplets. The error bar for each plot indicates the standard deviation.

Fig. 7 presents a comparison between the experimentally obtained freezing times and those calculated using Eq. (14) by substituting the maximum heat transfer coefficient. Given the stochastic nature of the freezing process, the experimental data underwent statistical analysis, and error bars were included to reflect the variability in the measurements. In this figure, the freezing time of cyclohexane is defined as the moment when the backlight is no longer transmitted (i.e., the droplet appears fully black) because of complete freezing. The freezing times derived from the experimental data agreed well with the model predictions within uncertainties, verifying the applicability of Eq. (14) for estimating the freezing time of a levitated droplet in an acoustic field. The fact that the experimental data fell below the model predictions is attributed to the uncertainties associated with the temperature difference and the determination of freezing time. While the model was effective in analyzing the onset of freezing, the complete freezing dynamics of levitated droplets require further investigation to clarify the evaporation-induced solidification process in acoustic levitation. Although this aspect falls beyond the scope of the present study, it will be addressed in future research.

## IV. CONCLUSIONS

The evaporation-induced dynamics of cyclohexane droplets in acoustic levitation were investigated. The experimental results confirmed that cyclohexane evaporates earlier than ethanol because of its higher saturation vapor pressure. Furthermore, cyclohexane droplets demonstrated that early evaporation leads to a transition to a supercooled state, with a surface temperature below the freezing point. This phenomenon is particularly evident in levitated droplets, which remain free from wall contact. Once the supercooling process was complete, the droplets rapidly froze. The experimental heat transfer coefficients of cyclohexane droplets agreed well with the model predictions based on the Ranz–Marshall correlations. For smaller droplet diameters, the experimental heat transfer coefficients were slightly higher than the model predictions owing to the complex convection driven by acoustic fields. Additionally, a model for predicting the onset of freezing time for levitated droplets considering the energy balance of evaporating droplets was proposed. These models suggest that heat transfer and freezing can be predicted to better characterize the complex phase change dynamics of levitated droplets in an acoustic field. However, the cooling dynamics were modeled through the rapid evaporation of the cyclohexane droplets to predict the onset of freezing time. Therefore, the discrepancy between the experiment and the prediction remains to be elucidated. Addressing this gap is a critical issue for future work, as freezing is a stochastic phenomenon. One potential approach involves the application of a non-equilibrium thermodynamic framework encompassing supercooling and ice nucleation.

While previous studies on acoustically levitated systems have primarily focused on evaporation behavior with external convective flows, this study offers new insights into the coupled heat and mass transport leading to evaporation-induced freezing. In particular, we experimentally demonstrated and modeled the onset of freezing in pure cyclohexane droplets under acoustic confinement, where both acoustic streaming and evaporative cooling contribute to the dynamics. The quantitative agreement between the measured and predicted heat transfer coefficients, along with the proposed energy-balance-based model for freezing onset, clarify non-equilibrium phase change phenomena in levitated droplets. These findings not only clarify evaporation-induced freezing dynamics but also pave the way for innovative developments in droplet-based technologies, such as microreactors.

## ACKNOWLEDGEMENTS

This work was supported by JSPS KAKENHI Grant Number 20H02070 and 23K17732.

## AUTHOR DECLARATIONS

### Conflict of Interest

The authors have no conflicts to disclose.

### Author Contributions

**Misaki Mitsuno:** Data curation (lead); Formal analysis (lead); Investigation (lead); Methodology (lead); Validation (lead); Visualization (lead); Writing – original draft (lead). **Xiao Ma:** Formal analysis (supporting); Investigation (supporting); Methodology (supporting); Writing – review & editing (supporting). **Koji Hasegawa:** Conceptualization (lead); Formal analysis (supporting); Investigation (supporting); Methodology (supporting); Validation (supporting); Visualization (supporting); Funding acquisition (lead); Project administration (lead); Supervision (lead); Writing – review & editing (lead).

## DATA AVAILABILITY

The data that support the findings of this study are available from the corresponding author upon reasonable request.